\documentclass{epl2}
\usepackage{amsmath}
\usepackage{amssymb}
\usepackage{subfig}
\usepackage[percent]{overpic} 

\newcommand{\be}{\begin{equation}}
\newcommand{\ee}{\end{equation}}
\newcommand{\bc}{\begin{center}}
\newcommand{\ec}{\end{center}}
\newcommand{\bi}{\begin{itemize}}
\newcommand{\ei}{\end{itemize}}
\newcommand{\ba}{\begin{eqnarray}}
\newcommand{\ea}{\end{eqnarray}}

\newcommand{\ignore}[1]{}

\title{Competition between surface relaxation and ballistic deposition
  models in scale free networks.}

\author{C. E. La Rocca\inst{1}\thanks{E-mail:
    \email{larocca@mdp.edu.ar}} \and P. A. Macri\inst{1} \and
  L. A. Braunstein\inst{1,2} }

\institute{ 
\inst{1} Investigaciones F\'isicas de Mar del Plata
  (IFIMAR)-Departamento de F\'isica, Facultad de Ciencias Exactas y
  Naturales, Universidad Nacional de Mar del Plata-CONICET, Funes
  3350, (7600) Mar del Plata, Argentina.\\ 
\inst{2} Center for Polymer
  Studies, Boston University, Boston, Massachusetts 02215, USA 
}

\shortauthor{C. E. La Rocca \etal}


\abstract{In this paper we study the scaling behavior of the
  fluctuations in the steady state $W_S$ with the system size $N$ for
  a surface growth process given by the competition between the
  surface relaxation (SRM) and the Ballistic Deposition (BD) models on
  degree uncorrelated Scale Free networks (SF), characterized by a
  degree distribution $P(k)\sim k^{-\lambda}$, where $k$ is the degree
  of a node. It is known that the fluctuations of the SRM model above
  the critical dimension ($d_c=2$) scales logarithmically with $N$ on
  euclidean lattices. However, Pastore y Piontti {\it et. al.}
  [A. L. Pastore y Piontti {\it et. al.}, Phys. Rev. E {\bf 76},
    046117 (2007)] found that the fluctuations of the SRM model in SF
  networks scale logarithmically with $N$ for $\lambda <3$ and as a
  constant for $\lambda \geq 3$. In this letter we found that for a
  pure ballistic deposition model on SF networks $W_S$ scales as a
  power law with an exponent that depends on $\lambda$. On the other
  hand when both processes are in competition, we find that there is a
  continuous crossover between a SRM behavior and a power law behavior
  due to the BD model that depends on the occurrence probability of
  each process and the system size. Interestingly, we find that a
  relaxation process contaminated by any small contribution of
  ballistic deposition will behave, for increasing system sizes, as a
  pure ballistic one. Our findings could be relevant when surface
  relaxation mechanisms are used to synchronize processes that evolve
  on top of complex networks.}

\begin{document}
\maketitle

In the last decade the study of complex networks received much
attention because many processes take place on top of these kinds of
structures. Here, we consider networks with scale free (SF) topologies
characterized by a degree distribution $ P(k) \sim k^{-\lambda}$,
where $k$ is the degree or number of connections that a node can have
with $k_{max} \ge k \ge k_{min}$, where $k_{max}$ and $k_{min}$ are
the maximal and minimal degree respectively, and $\lambda$ measures
the heterogeneity of the distribution
\cite{Barabasi_sf,nota}. Historically, the research on complex
networks was mainly focused on how the topology affects processes that
evolve on top of them, such as epidemic spreading
\cite{Pastorras_PRL_2001}, traffic flow
\cite{Lopez_transport,zhenhua}, cascading failures
\cite{Motter_prl,Sergey} and synchronization
\cite{Jost-prl,Korniss07,Kornissprl}.  Recently, it was shown that the
scaling behavior of processes, such as synchronization and jamming
\cite{criscorr,anacorr}, are fully governed by the topology of the
underlying network. Synchronization processes are very important in
many real situations such as supply-chain networks based on electronic
transactions \cite{Nagurney}, neuronal networks \cite{wang} and
financial transaction between traders \cite{saavedra}. One of the most
successful attempts to model synchronization phenomenon is to map them
into a non equilibrium surface growth \cite{Kornissprl,Kornissrev}
where the relevant magnitude that represents the departure from
synchronization is the dispersion or fluctuations of some scalar field
$h$ over the nodes of a network given by $W = \left\{1/N\sum_{i=1}^N
(h_i-\langle h \rangle)^2\right\}^{1/2}$, where $h_i$ represents the
scalar on node $i$, $\langle h \rangle$ is the mean value, $N$ is the
system size and $\{ . \}$ denotes an average over network
configurations. Synchronization problems deal with the optimization of
the fluctuations in the steady state, being the system optimally
synchronized when those fluctuations are minimized. On the other hand
a system is called scalable if their fluctuations do not depend, or
depend weakly, on the system size.

Many dynamic synchronization processes can be described by simple
growth models. One of the most successful models of surface growth
with relaxation is the Family model \cite{family} (SRM) that
represents very well film growth either by vapor or chemical
deposition. Other very important model is the Ballistic Deposition
(BD) \cite{Meakin} that represents sedimentation processes such as low
thin-film grow at low temperatures \cite{Meakin2}. However, more
complex processes cannot be described by a single growth model and are
better represented as the competition between two or more simple
mechanisms. The competition between models on euclidean lattices
\cite{Pellegrini,Horowitz,Chame,Muraca} was studied by few
researchers, in spite of the fact that it is more realistic in
describing grow processes on real materials. Pellegrini {\it et. al.}
\cite{Pellegrini} studied the competition between the SRM model with
probability $p$ and the BD model with probability $1-p$ on euclidean
lattices and found a crossover between a logarithmic and a power law
behavior of the fluctuations in the steady state $W_S$ with $N$ that
depends on $p$ for dimensions $d \ge 2$. Note that $d=2$ is the
critical dimension $d_c$ of the SRM model above which the behavior of
$W_S$ takes the mean field value, with $W_S \sim \ln N$. Pastore y
Piontti {\it et. al.}  \cite{anita} studied the SRM model on degree
uncorrelated SF networks and found that
\begin{eqnarray}\label{srmmodelo}
W_S\sim \begin{cases} \begin{matrix} \ln N & \mbox{for}\ \lambda < 3\ ,\\
const. & \mbox{for}\ \lambda \geq 3 \ ;\end{matrix} \end{cases}
\end{eqnarray}
a very different scaling behavior than the one obtained on euclidean
lattices above $d_c$, eventhough complex networks can be thought as
high dimensional systems with $d \gg d_c$. These results were
theoretically confirmed by La Rocca {\it et. al.}  \cite{cris} who
derived the evolution equation of this model on complex
networks. However, up to our knowledge, neither the BD model nor the
competition between different processes on complex networks was ever
reported. Usually the SRM model is associated with diffusion between
first neighbors of a network, such as in load balance problems in
parallel processors, where the load are tasks to solve. If the load
difference between processors is large, nodes with less load must wait
for the more loaded nodes to finish their tasks. Thus it is logical to
send all the new task to the less loaded node to prevent the
desinchronization of the system.  Then sending new tasks to the less
loaded processors could be represented by the BD model.

In this letter we study the scaling behavior of the fluctuations in
the steady state $W_S$ with the system size $N$ for the BD model, and
the competition between the SRM and the BD models on degree
uncorrelated SF networks.

In our simulations, at each time step a node $i$ is chosen with
probability $1/N$. Then, the SRM rules are applied with probability
$p$ and the BD rules with probability $1-p$. Denoting by $v_i$ the
nearest neighbor nodes of $i$, the evolution rules for the SRM model
are
\begin{eqnarray}
\begin{array}{ll}
(1)\ \mbox{if}\ h_i \leq h_j \forall j \in v_i \Rightarrow h_i= h_i+1,\ \mbox{else}\ ;\nonumber\\
(2)\ \mbox{if}\ h_j < h_n \forall n \not= j \in v_i \Rightarrow h_j=h_j+1\ .
\end{array}\nonumber
\end{eqnarray}
Meanwhile for the BD model the evolution rules are
\begin{eqnarray} 
\begin{array}{ll}
(1)\ \mbox{if}\ h_i \geq h_j \forall j \in v_i \Rightarrow h_i= h_i+1,\ \mbox{else}\ ,\\
(2)\ \mbox{if}\ h_n \geq h_j \forall j \not= n \in v_i \Rightarrow h_i = h_n\ .\nonumber
\end{array}\nonumber
\end{eqnarray}
As initial condition we choose $\{h_i\}=0$ for $i=1,...,N$ and the
time step increases by $1/N$. The SF networks with different values of
$\lambda$ were constructed using the configurational model
\cite{Molloy}. 

\begin{figure}[h]
\vspace{0.6cm}
\centering
\includegraphics[scale=0.35]{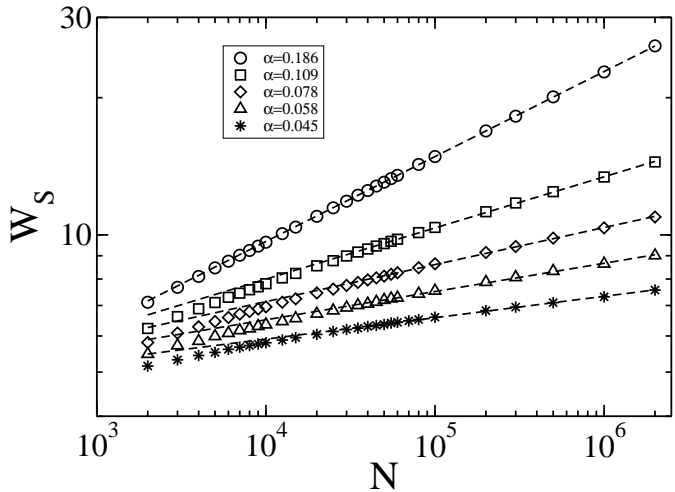}
\caption{Fluctuations $W_S$ as function of $N$ in log-log scale for
  pure BD model ($p=0$) in SF networks for $\lambda=2.5$ ($\bigcirc$),
  $2.8$ ($\Box$), $3.0$ ($\diamond$), $3.2$ ($\bigtriangleup$) and
  $3.5$ ($*$). The dashed lines correspond to a power law fitting for
  large system sizes $N$ with $W_S\sim N^\alpha$. \label{fig.1}}
\vspace{0.4cm}
\end{figure}

\begin{figure}[h]
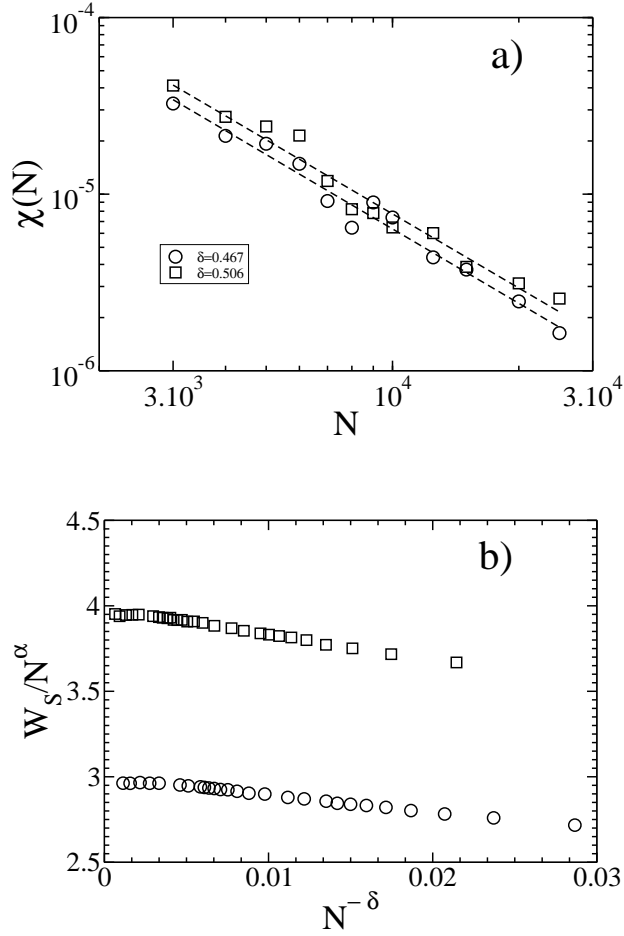

\centering
\vspace{0.5cm}
  \begin{overpic}[scale=0.31]{grafico1.eps}
  \end{overpic}\vspace{1cm}
  \begin{overpic}[scale=0.3]{grafico2.eps}
  \end{overpic}\vspace{0.5cm}
\caption{a) Corrections to the scaling $\chi(N)$ as function of $N$ in
  log-log scale for $\lambda=2.8$ ($\bigcirc$) and $3.5$ ($\Box$). The
  dashed lines correspond to the power law fitting from we obtain
  $\delta$. b) Linear-linear plot of $W_S/N^\alpha$ as function of
  $N^{-\delta}$ for the same values of $\lambda$ than in
  a). \label{fig.2}}
\end{figure}

In Fig. \ref{fig.1} we show the fluctuations $W_S$ as a function of
$N$ for the BD model ($p=0$) in log-log scale for different values of
$\lambda$. We can see that for all the values of $\lambda$ there
exists a power law behavior denoted by the dashed lines that represent
the fitting of the data with $W_S\sim N^\alpha$ for large system
sizes, where the exponent $\alpha$ depends on $\lambda$. Clearly, we
can observe that this behavior is very different than the one found
for the SRM model on SF networks (see Eq. (\ref{srmmodelo})). The
power law behavior of the fluctuations in the steady state of the BD
model seems to be an inherent property of the model regardless of the
topology of the complex network over which the process evolves,
although the exponent of the power law depends on the heterogeneity of
the network. We find that as the heterogeneity of the network
decreases, above $\lambda=2.5$, finite size effects become more
relevant modifying the power law behavior found for large $N$. Thus,
we propose that

\begin{equation}\label{propec}
W_S\sim N^{\alpha}(1+AN^{-\delta})\ ,
\end{equation}
where the term $N^{-\delta}$ is the correction to finite size scaling
with $\delta >0$ and $A$ a constant. We can measure the finite size
effects taking the derivative of $W_S/N^\alpha$ with respect to $N$
\begin{equation}
\chi(N) =\frac{\partial}{\partial N}\left(\frac{W_S}{N^\alpha}\right)\sim N^{-(\delta+1)}\ .\nonumber
\end{equation}
In Fig. \ref{fig.2} a) we plot $\chi(N)$ as a function of $N$ up to
$N=25000$, for two values of $\lambda >2.5$. The dashed lines
represent the fitting with a power law from where we obtain the values
of $\delta$. In order to show that our proposed scaling (see
Eq. (\ref{propec})) is correct in Fig. \ref{fig.2} b) we plot
$W_S/N^\alpha$ as function of $N^{-\delta}$. We can see that for large
system sizes $W_S/N^\alpha$ does not depend on $N$, but as $N$
decreases we observe a linear behavior, which shows the agreement
between the proposed corrections to scaling and the simulations.

\begin{figure}[h]
\vspace{0.7cm}
\centering
\includegraphics[scale=0.33]{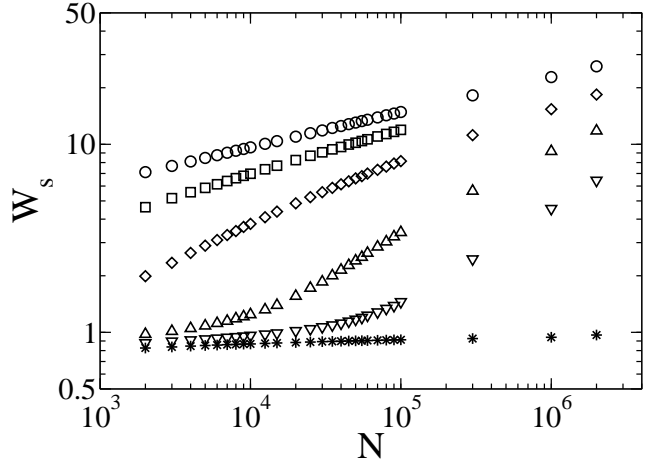}
\caption{Fluctuations $W_S$ as function of $N$ in log-log scale for SF
  networks with $\lambda=2.5$, $k_{min}=2$, for $p=0$ ($\bigcirc$),
  $0.6$ ($\Box$), $0.8$ ($\diamond$), $0.9$ ($\bigtriangleup$), $0.95$
  ($\bigtriangledown$), and $1$ ($\star$).\label{fig.3}}
\end{figure}

Next, we study the competition between the SRM and the BD models,
i.e., the scaling behavior of $W_S$ with $N$ when we increase the
parameter $p$. Although the conclusions reached from the competition
between the two models are qualitatively the same for all the values
of $\lambda$, we will consider only the case $\lambda=2.5$ because the
effects found are more pronounced in this case and because for this
value of $\lambda$ finite size effects are negligible as we mention
above. In Fig. \ref{fig.3} we plot $W_S$ as function of $N$ in log-log
scale for different values of $p$. We can see that for $p=0$ there is
a pure power law with exponent $0.186$ (see Fig \ref{fig.1}) and for
$p=1$ the scaling behavior of $W_s$ with $N$ corresponds to a
logarithm as expected for $\lambda <3$ \cite{anita, cris}. For values
$0 < p < 1$ we find that there is a crossover between both
regimes. The corresponding curves become steeper as $p$ increases
leading to expect that they will cross at some point. However, as the
system size is increased we observe that the curves change their
slope, avoiding the crossing between them. This is can be expected
since the fluctuations in the competition with the SRM model, which
tend to smooth out the interface via diffusion, cannot overcome the
fluctuations of the pure BD model. In fact, we expect that for large
sizes ($N \to \infty$), the behavior of all the curves with $p<1$ will
tend to follow a power law with the same exponent that for the case
$p=0$. However from our data we cannot show our expectation even for
the system sizes as big as the ones that we have achieved ($N \sim
10^6$). Note that each point in Figs. \ref{fig.1} and \ref{fig.3}
corresponds to $10^4$ realizations of the networks for fix $N$ where
each value of $W_S$ was obtained by a linear fitting on the steady
state. Thus to reach bigger system sizes is very time consuming and
goes beyond our computational capabilities. For this reason from the
results shown until here we can only conclude that in the competition
between both models, even for $p$ close to one, the BD dominates the
behavior of the fluctuations.

\begin{figure}[h]
\vspace{0.1cm}
\centering
\includegraphics[scale=0.8]{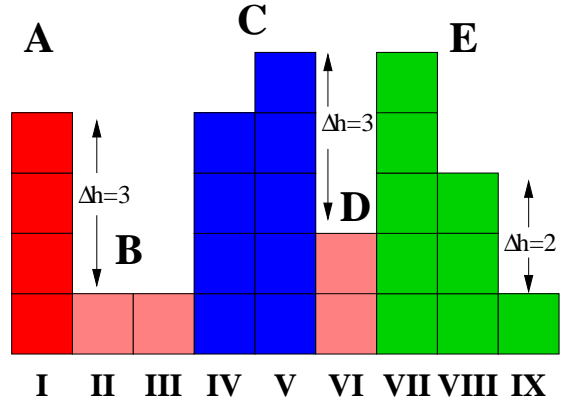}
\caption{Schematic of the formation of steps for a one-dimensional
  lattice where we use the threshold $\Delta h=2$. The steps,
  represented by different colors and capital letters, are bounded by
  nodes whose height difference exceeds the value of $2$. The
  numbering in the horizontal axis represents the node number and the
  vertical axis is the height of each node.\label{fig.4}}
\end{figure}

In the competition, the BD model is responsible of generating bigger
height differences (jumps) between neighboring nodes than the SRM
model, which tends to smooth out the interface by diffusion (see
Fig. \ref{fig.3}). As a consequence, the big jumps that produce the BD
model increase the fluctuations of the system. As $N$ increase, the
number of jumps also increases and as a consequence the BD model
dominates the behavior of the system, even for $p$ close to one. In
order to verify our assumption we introduce the definition of
``step''. Steps are landing structures with edges generated by the BD
process and flat regions smoothed by the SRM process. Since $W_S$
represents the dispersion of the heights around their mean value, we
can assume that most of the nodes of the network have a height $h
\approx \langle h \rangle \pm m W_S$, where $m$ is a positive integer
number that determines the degree of confidence for a Gaussian
distribution, i.e., $m=1$ correspond to $66\%$ of confidence and $m=3$
to $99\%$. We verified that the distribution of heights is a Gaussian
distribution for any value of $p$. This implies that in average the
absolute value of the difference of heights between any two pair of
neighboring nodes is at most
\begin{equation}
\Delta h \approx \sqrt 2 m W_S \ .\nonumber
\end{equation}
Then, if we denote the fluctuations for the pure SRM model ($p=1$) as
$W_S^{SRM}$, we define a step as an almost flat connected region of
the interface where each pair of neighboring nodes has $\Delta h \leq
\sqrt 2 m W_S^{SRM}$. Thus, computing how many steps each model
generates and how this magnitude changes with $p$, we can measure the
influence of the BD process \cite{stepsref}. In Fig. \ref{fig.4} we
show for $\Delta h >2$ a schematic plot for the particular case of a
one-dimensional lattice where each step is marked with a different
color and labeled by a letter. For example, the step C consist of
nodes IV and V that have $\Delta h <2$ between them, and is limited by
nodes III and VI because their height difference is greater than
$2$. Note that a step can be formed by a single node, as in the cases
A and D. Thus a step is a soft region of the interface bounded by two
borders represented by nodes with an abrupt jump in their height
differences ($\Delta h > \sqrt 2 m W_S^{SRM}$).

\begin{figure}[h]
\vspace{0.5cm}
\centering
\includegraphics[scale=0.33]{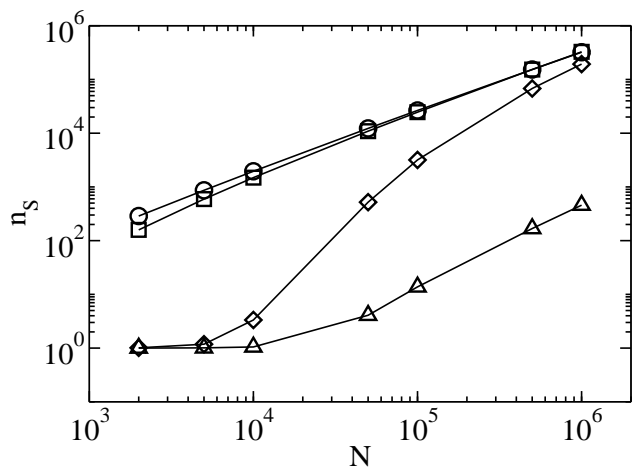}
\caption{Average number of steps $n_S$ as a function of $N$ in the
  surface for $m=3$ in log-log scale for SF networks with
  $\lambda=2.5$ and $p=0$ ($\bigcirc$), $0.6$ ($\Box$), $0.9$
  ($\diamond$) and $1$ ($\bigtriangleup$). The lines are used as
  guides for the eyes.\label{fig.5}}
\end{figure}

In Fig. \ref{fig.5} we plot in log-log scale the average number of
steps $n_S$ as a function of $N$ for $p=0$, $0.6$, $0.9$ and $1$, for
$m=3$ and taking $W_S^{SRM} \approx 1$ (see Fig. \ref{fig.3}). We can
see that for $p=1$, $n_S$ is much smaller than for $p=0$ for all the
values of $N$. For large $N$ and $0<p<1$ the curves of $n_S$ try to
reach the value of $p=0$, which means that the BD is the dominating
mechanism (see Fig. \ref{fig.3}). Thus even for values of $p$ very
close to $1$, corresponding to pure SRM, as $N$ increases more steps
appear and the BD model ends dominating the behavior of the
fluctuations in the steady state. This explain the crossover behavior
of $W_S$ with $N$ for all the values of $p<1$. Thus a relaxation
process driven by diffusion competing with a small contribution of the
ballistic deposition will behave as a pure ballistic one for large
$N$. Then, if we are interested in reducing the fluctuations in order
to improve the synchronization of the system when these two processes
are competing, we must increase the diffusion and reduce the system
size. Our findings could have a great impact when surface relaxation
mechanisms are used to synchronize system on complex networks because
a small contribution of the BD process reduces the synchronization and
as a consequence the scalability of the system with the system size
worsens \cite{lam3}.

In summary, we studied the scaling behavior of the fluctuations on the
steady state for the ballistic deposition model in degree uncorrelated
SF networks and found that $W_S\sim N^{\alpha}$ where the exponent
$\alpha$ increases with the heterogeneity of the SF networks, i.e., as
$\lambda$ decreases. Our results suggest that the power law behavior
of $W_S$ is an inherent property of this process found also in
euclidean lattices. We also studied the scaling of $W_S$ with $N$ for
the competition between the SRM and the BD models and found that there
exists a crossover between different behaviors as we increase $p$. For
large $N$, we found that for $p<1$ the values of $W_S$ approaches to
the BD one, which means that the BD model always dominates the
behavior of the fluctuations in the steady state. This was shown using
the definition of step, where the average number of steps measure the
influence of the BD process. We found that the crossover between the
SRM and the BD corresponds to an abrupt increase in the average number
of steps.

In a forthcoming work we will study analytically the BD model and the
competition between the BD and the SRM models deriving the evolution
equations for the heights from where we will be able to derive the
scaling behavior of $W_S$ for all the values of $p$.

\acknowledgments We thank UNMdP and FONCyT (PICT 0293/08) for
financial support.


\begin{thebibliography}{0}
\bibitem{stepsref} While quantitative results depend on the value of
  $m$, the qualitative results do not, since the number of steps will
  change depending on $m$ but the crossover between the two models
  will be clearly presented for any $m \ge 1$.
\bibitem{lam3} The results obtained for $\lambda \ge 3$ are very
  similar to the ones found for $\lambda=2.5$, although the crossover
  between the two behaviors now is between a constant, consequence of
  the SRM model for $\lambda \ge 3$ (see Eq. (\ref{srmmodelo})), and a
  power law.
\end{thebibliography}
\end{document}